\documentstyle[prl,aps,epsfig,twocolumn]{revtex}

\begin{document}

\draft

\twocolumn[\hsize\textwidth\columnwidth\hsize
\csname@twocolumnfalse\endcsname
\title{Quantum Memory Based on $\Lambda$-Atoms Ensemble
with Two-Photon Resonance EIT}
\author{Y. Li and C. P. Sun$^{a,b}$}
\address{Institute of Theoretical Physics, Chinese
Academy of Sciences, Beijing, 100080, China}
 \maketitle

\begin{abstract}
We study the $\Lambda$-atoms ensemble based quantum memory for the
storage of the quantum information carried by a probe light field.
Two atomic Rabi transitions of the ensemble are coupled to the
quantum probe field and classical control field respectively with a
same detuning. Making use of the hidden symmetry analysis developed
recently for the on-resonance EIT case (Sun, Li, and Liu, Phys.
Rev. Lett. 91, 147903 (2003)), we show that the dark states and
dark-state polaritons can still exist for the case of two-photon
resonance EIT. Starting from these dark states we construct a
complete class of eigen-states of the total system. A explicit form
of the adiabatic condition is also given in order to achieve the
memory and retrieve of quantum information.
\end{abstract}
  ]

With great interest and quick development in quantum information
science \cite{q-infor}, the implementation of quantum memory
becomes one of the particular challenges to quest a realistic
system transporting or communicating quantum states between
different nodes of quantum networks. Quantum optical systems with
atoms appear to be very attractive since photons are ideal carriers
of quantum information with very fast velocity; and the atoms
represent long-lived storage and reliable processing units.

A well known quantum optical system is of the electromagnetically
induced transparency (EIT) \cite{1}, which can be used to make a
resonant, opaque medium transparent. The essential property of EIT
is induced by atomic coherence and quantum interference. The
discovery of EIT has led to the occurrence of new effects and new
techniques including ultraslow light pulse propagation \cite{2,3}
and the light signal storage \cite{4,Fl00-OptCom} in atomic vapor.
A following idea is then how to use the EIT system to transport and
communicate the quantum state between photons and atoms.

Conventionally the EIT system consists of a vapor of 3-level atoms
with two classic optical fields (the probe and control fields)
being one-photon on-resonance with the relevant atomic transitions
\cite{1,2,Scullybook}. Later, it is noticed that such an
on-resonance EIT is not a prerequisite for achieving significant
group velocity reduction. The EIT phenomenon can also occur when
the frequency difference between the probe and control fields
matches the two-photon transition between the two lower states of
the $\Lambda $-type atoms \cite{Deng01,Deng02,Lukin-rmp} (for this
case, it is called two-photon resonance EIT).

In order to implement the quantum memory and to transport the quantum states
between photons and atomic ensemble, recently some people \cite%
{4,Fl00-OptCom,Lukin00-ent,Fl00-pol,Andre02,Fl02-DSP} have replaced
the classical probe laser field by a weak quantum light field in
the on-resonance EIT system with atomic collective excitations.
Then, by adiabatically changing the coupling strength of the
classic control field, they have demonstrated the possibility of
coherently controlling the propagation of the quantum light pulses
via the dark sates and dark-state polaritons. Most recently, to
avoid the spatial-motion induced decoherence, we have considered an
on-resonance EIT system of "atomic crystal" with each atom fixed on
a lattice site \cite{Sun01}. With discovery of the hidden dynamic
symmetry, we have shown that such a system is a robust quantum
memory to transport the quantum states between photons and atomic
ensemble.

Since most works about the memory of quantum probe light field
within an atomic ensemble are based on the on-resonance EIT with
atomic collective excitations \cite{Lukin00-ent,Fl00-pol,Sun01}, we
want to study a system under the case of two-photon resonance EIT
with the method of atomic collective excitations. In a former paper
\cite{li and sun} we have calculated the susceptibility and group
velocity of the probe field under two-photon resonance. Our results
show that the EIT phenomenon exists indeed and an ultraslow group
velocity can be obtained. In this work, we study how the excitonic
system under two-photon resonance EIT serves as a robust quantum
memory.

We consider an atomic ensemble consisted of $N$ 3-level atoms of $\Lambda $%
-type, which are coupled to two single-mode optical fields as shown in Fig.
1. The atomic levels are labelled as the ground state $|b\rangle $, the
excited state $|a\rangle $ and the meta-stable state $|c\rangle $. The
atomic transition $|a\rangle \leftrightarrow |b\rangle $ with energy level
difference $\omega _{ab}$ $=\omega _{a}-\omega _{b}$ is coupled to a quantum
probe light field of frequency $\omega $ with the coupling coefficient $g$
and the detuning $\Delta _{p}=\omega _{ab}-\omega $, while the atomic
transition $|a\rangle \leftrightarrow |c\rangle $ energy level difference $%
\omega _{ac}$ driven by a classical control field of frequency $v$ with the
Rabi-frequency $\Omega (t)$ and the detuning $\Delta _{c}=\omega _{ac}-\nu $%
. For simplicity, the coupling coefficients $g$ and $\Omega $ are real and
assumed to be identical for all the atoms in the ensemble. Under the
two-photon resonance condition, that is, $\Delta _{p}\equiv \Delta _{c}$,
the interaction Hamiltonian of total system can be written in the
interaction picture as ($\hbar =1$)
\begin{equation}
H_{I}=\Delta _{c}S+(g\sqrt{N}aA^{\dagger }+\Omega T_{+}+h.c.)
\end{equation}%
in terms of the collective quasi-spin operators

\begin{eqnarray}
S &=&\sum_{j=1}^{N}\sigma _{aa}^{(j)},  \nonumber \\
\ A^{\dagger } &=&\frac{1}{\sqrt{N}}\sum_{j=1}^{N}\sigma _{ab}^{(j)}, \\
T_{+} &=&\sum_{j=1}^{N}\sigma _{ac}^{(j)}.  \nonumber
\end{eqnarray}%
Here $\sigma _{\mu \nu }^{(j)}=|\mu \rangle _{jj}\langle \nu |$ is the flip
operator of the $j$-th atom from state $|\mu \rangle _{j}$ to $|\nu \rangle
_{j}$ ($\mu ,\nu =a,b,c$); and $a^{\dagger }$ ($a$) is the creation
(annihilation) operator of the probe field. In the large $N$ and low atomic
excitation limit with only a few atoms occupy states $|a\rangle $ or $%
|c\rangle $ \cite{q-defor}, the quasi-spin wave excitations of the atoms
behave as bosons since in this case they satisfy the bosonic commutation
relation $[A,A^{\dagger }]=1$.

\begin{figure}[h]
\hspace{24pt}\includegraphics[width=6cm,height=6cm]{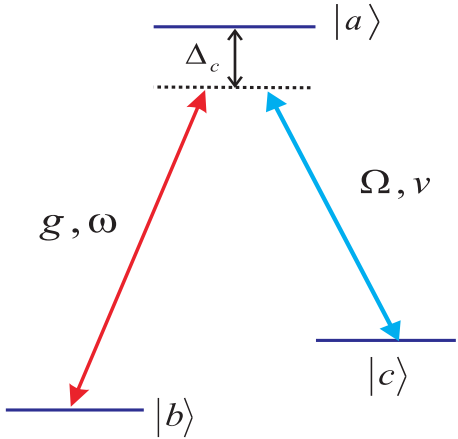}
\caption{The probe and control optical fields are respectively coupled to
two atomic transitions with the same detuning $\Delta_ {c} $. That is, such
a system consisted of 3-level $\Lambda$-atoms ensemble and two optical
fields satisfies the two-photon resonance EIT condition.}
\end{figure}

We note that the above Hamiltonian is expressed in terms of the collective
dynamic variables $S$, $A$, $A^{\dagger }$, $T_{+}$, and $%
T_{-}=(T_{+})^{\dagger }$. To properly describe the cooperative motion of
the atomic ensemble stimulated by the probe and control fields, we consider
the closed Lie algebra generated by $A$, $A^{\dagger }$, $T_{+}$, and $T_{-}$%
. To this end a new pair of collective excitation operators
\begin{equation}
C=\frac{1}{\sqrt{N}}\sum_{j=1}^{N}\sigma _{bc}^{(j)},\ \ C^{\dagger
}=(C)^{\dagger }
\end{equation}%
are introduced here to form a closed algebra. In the large $N$ and low
excitation limit, the corresponding collective excitations also behave as
bosons since they satisfy the bosonic commutation relation $[C,C^{\dagger
}]=1$. These quasi-spin collective excitations are independent of each other
in the same limit because of the vanishing commutation relations%
\begin{equation}
\lbrack A,C]=0,\text{ }[A,C^{\dagger }]=-T_{-}/N\longrightarrow 0
\end{equation}%
by a straightforward calculation. It's noted that $S$ is a Hermitian
operator and has the commutation relations%
\begin{eqnarray}
\lbrack S,A] &=&-A,  \nonumber \\
\lbrack S,A^{\dagger }] &=&A^{\dagger },  \nonumber \\
\lbrack S,C] &=&[S,C^{\dagger }]=0,  \label{s-commu} \\
\lbrack S,T_{\pm }] &=&\pm T_{\pm }.  \nonumber
\end{eqnarray}%
Moreover, it's easy to prove the following basic commutation relations%
\begin{eqnarray}
\lbrack T_{-},C] &=&-A,\text{ }[T_{-},C^{\dagger }]=0,  \nonumber \\
\lbrack T_{-},A] &=&0,\text{ }[T_{-},A^{\dagger }]=C^{\dagger }.
\label{T-A-C-commu}
\end{eqnarray}%
The above relations define a dynamic symmetry hidden in our dressed atomic
ensemble described by the semi-direct-product algebra $SU(2)\overline{%
\otimes }h_{2}$ containing the algebra $SU(2)$ (which is generated by $T_{-}
$, $T_{+}$, and%
\begin{equation}
T_{3}=\sum_{j=1}^{N}(\sigma _{aa}^{(j)}-\sigma _{cc}^{(j)})/2
\end{equation}
as the third generator) and the algebra $h_{2}$ (generated by $A$, $%
A^{\dagger }$, $C$, and $C^{\dagger }$), since it follows that%
\begin{equation}
\lbrack SU(2),h_{2}]\subset h_{2}.
\end{equation}

Actually, a similar semi-direct product algebra
$SU(2)\overline{\otimes }h_{2}$ has been found by us \cite{Sun01}
in an "atomic crystal" under the case of the one-photon resonance
EIT. In that paper, by means of this algebra and the spectral
generating algebra method, we have obtained a complete class of
eigen-states of the total system consisted of the atoms and optical
fields. And a explicit form of adiabatic passage has been given to
implement the quantum information memory and retrieve between the
type of photons and the type of atomic collective excitations. In
what follows, we will study how the system shown in this work can
serve as a robust quantum memory.

We can define a polariton operator
\begin{equation}
D=a\cos \theta -C\sin \theta ,  \label{ddd}
\end{equation}%
where $\theta (t)$ satisfies%
\begin{equation}
\tan \theta (t)=\frac{g\sqrt{N}}{\Omega (t)}.
\end{equation}%
This polariton operator that mixes the optical field and the atomic
collective excitations behaves a boson in the large $N$ and low excitation
limit since%
\begin{equation}
\lbrack D,D^{\dagger }]=1.
\end{equation}%
Thus the Heisenberg-Weyl group $h$ generated by $D$ and $D^{\dagger }$ is a
symmetry group of the exciton-photon system. We introduce the state $|{\bf 0}%
\rangle =|0\rangle _{p}\otimes |b^{N}\rangle $ where $|0\rangle _{p}$ is the
vacuum of the electromagnetic field and $|b^{N}\rangle $ $%
=|b_{1}b_{2}...b_{N}\rangle $ denotes all the $N$ atoms staying in the
ground states. The relations%
\begin{equation}
H_{I}|{\bf 0}\rangle =0,\text{ }[D,H_{I}]=0
\end{equation}%
hint to us that a degenerate class of eigen-states of $H_{I}$ with
zero eigen-value can be constructed naturally as follows:
\begin{equation}
|d_{n}\rangle =[n!]^{-1/2}D^{\dagger n}|{\bf 0}\rangle ,\text{ }%
n=0,1,2,\cdots .  \label{dark}
\end{equation}%
Using the Eq. (\ref{ddd}), we can expand $|d_{n}\rangle $ as%
\begin{eqnarray}
|d_{n}\rangle &=&\sum_{m=0}^{n}(-1)^{m}\sqrt{\frac{n!}{m!(n-m)!}}\cos
^{n-m}\theta \sin ^{m}\theta  \nonumber \\
& \times &|{\bf c}^{m}\rangle \otimes |n-m\rangle _{p},\text{ }n
=0,1,2,\cdots .
\end{eqnarray}%
where%
\begin{equation}
|{\bf c}{^{m}\rangle =}[m!]^{-1/2}C^{\dagger m}|b^{N}\rangle
\end{equation}%
represents there are $m$ $C$-mode excitations in the atomic
ensemble. Let us draw the energy level graphic as Fig. 2. It can be
observed from Fig. 2 that the dark state $|d_{n}\rangle $ is a
linear combination consisted of the terms $|b^{N}\rangle \otimes
|n\rangle _{p}$, $|{\bf c}\rangle \otimes |n-1\rangle _{p}$, $...$,
$|{\bf c}^{n}\rangle \otimes |0\rangle _{p}$. In this work, we
ignore all the atomic decays. Generally the decay related to the
atomic excited state $|a\rangle $ is much larger than the decay
related to the meta-stable state $|c\rangle $. However, the atomic
parts of all the terms in
the dark state only contain the single-atom lower states $|b\rangle $ and $%
|c\rangle $ so that the dark state is robust even if the physical decays are
considered.

\begin{figure}[h]
\hspace{24pt}\includegraphics[width=6cm,height=5cm]{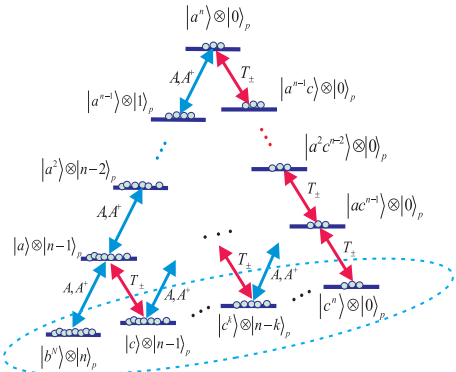}
\caption{The schematics of the states in the two-photon resonance
system with atomic collective excitations, where $|{\bf a}^{l}{\bf
c}{^{m}\rangle =}[l!m!]^{-1/2}A^{\dagger l}C^{\dagger
m}|b^{N}\rangle $ denotes there are $l$ $A$-mode and $m$ $C$-mode
excitations in the atomic ensemble. These states in the figure are
connected by the interaction of the operators $A$, $A^{\dagger}$,
and $T_{\pm}$ (also the action of the related optical fields). The
dark state $|d_{n}\rangle $ is a superposition of
the states $|b^N \rangle \otimes|n\rangle _{p}$, $|{\bf c}\rangle \otimes |n-1\rangle _{p}$, $...$, $|{\bf c}%
^{n}\rangle \otimes |0\rangle _{p}$ in the dashing envelop.}
\end{figure}

Physically, the above dressed state is cancelled by the interaction
Hamiltonian, thus it is called a dark state and $D$ is called a dark-state
polariton (DSP). The DSP traps the electromagnetic radiation from the
excited state due to the quantum interference cancelling. For the case with
an ensemble of free moving atoms, the similar DSP was obtained in Refs. \cite%
{Fl00-OptCom,Lukin00-ent,Fl00-pol,Sun01} to clarify the physics of
the state-preserving slow light propagation in EIT associated with
the existence of collective atomic excitations.

Now starting from these dark states $|d_{n}\rangle $, we can use the
spectrum generating algebra method to build other eigenstates for the total
system. To this end we introduce the bright-state polariton operator
\begin{equation}
B=a\sin \theta +C\cos \theta .
\end{equation}%
It is obvious that%
\begin{eqnarray}
\lbrack B,B^{\dagger }] &=&1,\text{ }  \nonumber \\
\lbrack D,B^{\dagger }] &=&[D,B]=0.
\end{eqnarray}%
Evidently $[A,B]=[A,B^{\dagger }]=0$ using the fact that $A$ commutes with $%
C $ and $C^{\dagger }$ in the large $N$ with low excitation limit.

It is straightforward to obtain the commutation relations%
\begin{eqnarray}
\lbrack H_{I},B^{\dagger }] &=&\varepsilon A^{\dagger },\text{ }  \nonumber
\\
\lbrack H_{I},A^{\dagger }] &=&\Delta _{c}A^{\dagger }+\varepsilon
B^{\dagger },  \label{dynamical}
\end{eqnarray}%
where $\varepsilon =\sqrt{g^{2}N+\Omega ^{2}}$. Notice that the system of
dynamical equations (\ref{dynamical}) are not simply as the same as that of
the case of on-resonance \cite{Sun01}. However, we can introduce the norm
mode variables $Q_{\pm }$:%
\begin{equation}
Q_{\pm }=\sqrt{\frac{\Theta \pm \Delta _{c}}{2\Theta }}A\pm \sqrt{\frac{%
\Theta \mp \Delta _{c}}{2\Theta }}B,
\end{equation}%
where%
\begin{equation}
\Theta =\sqrt{\Delta _{c}^{2}+4\varepsilon ^{2}}.
\end{equation}%
What is crucial for our purpose is the commutation relations%
\begin{equation}
\lbrack H_{I},Q_{\pm }^{\dagger }]=e_{\pm }Q_{\pm }^{\dagger },  \label{Q}
\end{equation}%
where%
\begin{equation}
e_{\pm }=\pm \sqrt{\frac{\Theta \pm \Delta _{c}}{\Theta \mp \Delta _{c}}}%
\varepsilon .  \nonumber
\end{equation}%
By introducing the norm mode transformation, it is almost the same as the
case of on-resonance EIT. Based on these commutation relations we can
construct the eigen-states
\begin{equation}
|e(m,k;n)\rangle =[m!k!]^{-1/2}Q_{+}^{\dagger m}Q_{-}^{\dagger
k}|d_{n}\rangle ,
\end{equation}%
as the dressed states of the total system. The corresponding eigen-values
are
\begin{equation}
E(m,k)=me_{+}-ke_{-},\text{ }m,k=0,1,2,\cdots .
\end{equation}

In the following discussion, we consider whether the dark states of
zero-eigen-value can work well as a quantum memory by the adiabatic
manipulation. This means that we should consider how the adiabatic condition
\cite{prd,zee}%
\begin{equation}
\left\vert \frac{\left\langle e(m,k;n)\right\vert \partial _{t}|d_{l}\rangle
}{E(m,k)-0}\right\vert \ll 1,
\end{equation}%
is satisfied for any $m,k,n,l=0,1,2,\cdots $. The eigen-values of these
instantaneous collective eigen-states are complicated and the corresponding
energy levels can cross each other (including the dark states) when
adiabatically varying the Rabi frequency $\Omega (t)$. Fortunately, among
all the terms $\left\langle e(m,k;n)\right\vert \partial _{t}|d_{l}\rangle $%
, only the terms $\left\langle e(0,1;l)\right\vert \partial
_{t}|d_{l}\rangle $ and $\left\langle e(1,0;l)\right\vert \partial
_{t}|d_{l}\rangle $ do not vanish. In general, we calculate

\begin{eqnarray}
&&\left\langle e(m,k;n)\right\vert \partial _{t}|d_{l}\rangle  \nonumber \\
&=&\frac{1}{\sqrt{m!k!n!l!}}\left\langle {\bf 0}\right\vert
Q_{+}^{m}Q_{-}^{k}D^{n}\partial _{t}D^{\dagger l}|{\bf 0}\rangle  \nonumber
\\
&=&\frac{l}{\sqrt{m!k!n!l!}}\left\langle {\bf 0}\right\vert
Q_{+}^{m}Q_{-}^{k}D^{n}D^{\dagger (l-1)}\partial _{t}D^{\dagger }|{\bf 0}%
\rangle  \nonumber \\
&=&\frac{-l\dot{\theta}}{\sqrt{m!k!n!l!}}\left\langle {\bf 0}\right\vert
Q_{+}^{m}Q_{-}^{k}D^{n}D^{\dagger (l-1)}B^{\dagger }|{\bf 0}\rangle
\nonumber \\
&=&\frac{-l!\dot{\theta}\delta _{n,l-1}}{\sqrt{m!k!n!l!}}\left\langle {\bf 0}%
\right\vert Q_{+}^{m}Q_{-}^{k}B^{\dagger }|{\bf 0}\rangle  \nonumber \\
&=&\sqrt{l}\dot{\theta}\delta _{n,l-1}[\delta _{m,0}\delta _{k,1}\sqrt{\frac{%
\Theta +\Delta _{c}}{2\Theta }}-\delta _{m,1}\delta _{k,0}\sqrt{\frac{\Theta
-\Delta _{c}}{2\Theta }}].  \label{condition}
\end{eqnarray}%
From the above results, we can readily obtain the adiabatic condition:%
\begin{equation}
\frac{g\sqrt{N}(\Theta +\left\vert \Delta _{c}\right\vert )}{\sqrt{\Theta
(\Theta -\left\vert \Delta _{c}\right\vert )}\varepsilon ^{3}}\left\vert
\dot{\Omega}\right\vert \ll 1  \label{adiabatic}
\end{equation}%
by means of%
\begin{equation}
\dot{\theta}=-g\sqrt{N}|\dot{\Omega}|/\varepsilon ^{2}.
\end{equation}%
When $\Delta _{c}=0$, the above equation (\ref{adiabatic}) will reduce to%
\begin{equation}
g\sqrt{N}|\dot{\Omega}|/\varepsilon ^{3}\ll 1,
\end{equation}%
which is just as the same as the adiabatic condition under on-resonance EIT
as shown in Ref. \cite{Sun01}. According to the Eq. (\ref{condition}), we
know that: the dark state $|d_{l}\rangle $ will not mix the states $%
|e(m,k;n)\rangle $ under the adiabatic condition (\ref{adiabatic}); and the
dark state $|d_{l}\rangle $ also can not mix the other degenerate state $%
|d_{l^{\prime }}\rangle $ according to the adiabatic condition of degenerate
states \cite{Sun01,zee}. This means when the initial state is%
\begin{equation}
|\Phi (0)\rangle =\sum_{n}c_{n}(0)|d_{n}(0)\rangle,
\end{equation}
the total system will follow the superposition of dark-state%
\begin{equation}
|\Phi (t)\rangle =\sum_{n}c_{n}(0)|d_{n}(t)\rangle
\end{equation}%
under the adiabatic evolution.

\begin{figure}[h]
\hspace{24pt}\includegraphics[width=6cm,height=6cm]{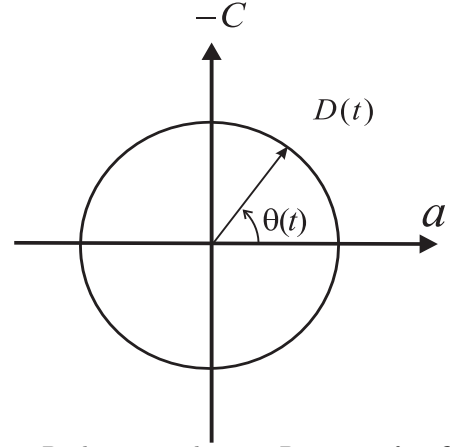}
\caption{Dark state polariton $D=a\cos \protect\theta -C\sin
\protect\theta $
is dependent on the parameter $\protect\theta (t)=\arctan \frac{g\protect%
\sqrt{N}}{\Omega (t)}$. If initially $\protect\theta (t_{0})=0$, $D(t_{0})=a$%
, and the state of the total system is described as $|\protect\psi %
(t_{0})\rangle $ $=|b^{N}{\rangle }\otimes |n\rangle _{p} $, then when $%
\protect\theta $ is adiabatically changed to $\protect\theta (t_{1})=\protect%
\pi /2$, one has $D(t_{1})=-C$ and the state is described as $|\protect\psi %
(t_{1})\rangle $ $=|c{^{n}\rangle }\otimes |0\rangle _{p}$.}
\end{figure}

Then we can implement the quantum information memory and retrieve between
the type of photons and the type of atomic collective excitations. As shown
in Fig. 2, we denote the dark state polariton $D(t)$ as a vector in the $a$-$%
C$ plane. Since the dark state is generated by $D^{\dagger }$ as
\begin{equation}
|d_{n}(t)\rangle =[n!]^{-1/2}D^{\dagger n}(t)|{\bf 0}\rangle ,
\end{equation}%
then if $\Omega (t)$ is changed adiabatically to make $\theta (t):$ $%
0\rightarrow \frac{\pi }{2}$, one have $D(t):$ $a\rightarrow -C\ $and $%
|d_{n}(t)\rangle $ will change from $|b^{N}{\rangle }\otimes |n\rangle _{p}$
to $|{\bf c}{^{n}\rangle }\otimes |0\rangle _{p}$. Generally, the initial
quantum state of the single-mode optical field is described by a density
matrix%
\begin{equation}
\rho _{p}=\sum_{n,m}\rho _{nm}|n\rangle _{pp}\langle m|,
\end{equation}%
the transfer process generates a quantum state of collective excitations
according to%
\begin{eqnarray}
&&|b^{N}{\rangle \langle }b^{N}|\otimes \sum_{n,m}\rho _{nm}|n\rangle
_{pp}\langle m|  \nonumber \\
&\rightarrow &\sum_{n,m}(-1)^{n+m}\rho _{nm}|{\bf c}{^{n}\rangle \langle
{\bf c}^{m}}|\otimes |0\rangle _{pp}{\langle 0}|.
\end{eqnarray}%
After an inverse adiabatic control to make $\theta :\frac{\pi }{2}%
\rightarrow 0$, the total quantum information will change from the type of
the atomic ensemble to the type of photons:%
\begin{eqnarray}
&&\sum_{n,m}(-1)^{n+m}\rho _{nm}|{\bf c}{^{n}\rangle \langle {\bf c}^{m}}%
|\otimes |0\rangle _{pp}{\langle 0}|  \nonumber \\
&\rightarrow &|b^{N}{\rangle \langle }b^{N}|\otimes \sum_{n,m}\rho
_{nm}|n\rangle _{pp}\langle m|.
\end{eqnarray}%
The quantum information can be adiabatically transferred from the optical
field to the atomic ensemble, and {\it vice versa}. Such two adiabatic
passages complement the "write" and "read" manipulation of quantum
information. That is to say, the quantum information can be memorized in
such an atomic ensemble.

In summary, we study the structure of the eigen-states and
eigen-values of the collective exciton-photon system under the
two-photon resonance EIT. Our results show that the dark states and
dark-state polaritons can still exist for the case of two-photon
resonance. We analyze in detail the possibility of a dark state
being staying the initial form under the adiabatic evolution of
systemic parameter $\Omega (t)$. A precise adiabatic condition is
presented in order to make sure that a dark state can not mix any
other eigen-states. Then, with the help of the dark states under
the two-photon resonance EIT system, we have described how one can
transfer or communicate the quantum states between the type of
photons and the type of atomic collective excitations by
adiabatically changing the Rabi frequency $\Omega (t)$ of the
classical control laser field. Our results show that the two-photon
resonance EIT system can be used as the same robust quantum memory
as that under the case of one-photon resonance EIT.

{\it This work is supported by the NSFC and the knowledge Innovation Program
(KIP) of the Chinese Academy of Sciences. It is also founded by the National
Fundamental Research Program of China with No. 001GB309310.}

\end{document}